\documentclass[aps,prb,superscriptaddress,showpacs,showkeys,twocolumn,amsmath,amssymb,reprint]{revtex4-1}

\usepackage{times}
\usepackage{color}
\usepackage{graphicx}
\usepackage{amsmath,amsbsy,amsfonts,mathrsfs}

\newcommand{\be}{\begin{equation}}
\newcommand{\ee}{\end{equation}}
\newcommand{\bea}{\begin{eqnarray}}
\newcommand{\eea}{\end{eqnarray}}
\newcommand{\bg}{\begin{figure}}
\newcommand{\eg}{\end{figure}}
\newcommand{\bi}{\begin{itemize}}
\newcommand{\ei}{\end{itemize}}

\usepackage[version=3]{mhchem} 
\usepackage{dcolumn}
\usepackage{textcomp}

\begin{document}
\bibliographystyle{apsrev}


\title[An Improved Descriptor of Cluster Stability. Application to Small Carbon Clusters]{An Improved Descriptor of Cluster Stability. Application to Small Carbon Clusters}

\author{\bf Jos\'e I. Mart\1nez}
\email{joseignacio.martinez@icmm.csic.es}
\address{Materials Science Factory, Dept. Surfaces, Coatings and Molecular Astrophysics, Institute of Material Science of Madrid (ICMM-CSIC), Sor Juana In\'es de la Cruz 3, ES-28049 Madrid, Spain}

\author{\bf Julio A. Alonso}
\address{Departamento de F\1sica Te\'orica, At\'omica y \'Optica, University of Valladolid, ES-47011 Valladolid, Spain}

\date{\today}

\begin{abstract}
The mass spectra of gas-phase clusters in cluster beams have a rich structure where the relative heights of the peaks compared to peaks corresponding to clusters of neighbor sizes reveal the stability of the clusters as a function of the size $N$. In an analysis of the published mass spectrum of carbon clusters cations $C_N^{+}$ with $N\leq$16 we have employed the most common descriptor of cluster stability, which is based on comparing the total energy of the cluster of size $N$ with the averaged energies of clusters with sizes $N$+1 and $N$-1. Those energies have been obtained from density functional calculations. The comparison between the stability function and the mass spectrum leaves some experimental features unexplained; in particular, the correlation with the detailed variation of the height of the mass peaks as a function of size $N$ is not satisfactory. We then propose a novel stability descriptor which improves matters substantially, in particular the correlation with the detailed variation of the height of the mass peaks. The new stability index is based on comparing the atom-evaporation energy of the cluster of size $N$ with the averaged atom-evaporation energies of clusters with sizes $N$+1 and $N$-1. The substantial improvement achieved is attributed to the fact that evaporation energies are quantities directly connected with the processes controllig the cluster abundances in the beam.
\end{abstract}

\maketitle

\section{Introduction}

Atomic and molecular clusters, and nanoparticles have enormous interest from both points of view of fundamental science and applications. The properties of clusters are often different from the properties of the macroscopic system, and vary with cluster size~\cite{AlonsoBook, AlonsoChemRev2000}. Changing the shape of nanoparticles has been used as a tool to modify their optical properties, for instance the color of the nanoparticles~\cite{PEREZJUSTE2004137}. Size and shape influence the catalytic properties of clusters and nanoparticles~\cite{HARUTA1997153}. This means that by playing with their size and shape, clusters and nanoparticles can be tailored for specific functionalities.
 
Different methods exist for producing clusters~\cite{AlonsoBook}. In one of the most common methods the solid material is first vaporized, and then the clusters grow by atom aggregation in a cluster beam. Clusters of different sizes are formed in this way, and the analysis of the abundance population as a function of cluster size in the molecular beam is made by mass spectrometry. The relative abundance of clusters with different sizes gives information on the relative stability of the clusters as a function of the cluster size $N$. In the mass spectrum of clusters $X_N$, where $X$ indicates the particular chemical element of interest, each cluster of size $N$ yields a peak, and the height of the peak measures the abundance of that size $N$ in the cluster beam. The clusters grow in the condensation chamber mainly by aggregation of new atoms, and this part of the process normally results in a smooth distribution of cluster sizes whose population decreases as $N$ increases. However, the clusters become hot as they grow due to the heat of condensation. Part of this energy can be liberated in collisions with the atoms of a carrier gas (Ar, for instance), but a substantial part of the condensation energy still remains in the cluster and induces the evaporation of one or more atoms. In this way, the size of the cluster shrinks and this evaporative effect leads to an enhancement of the population of the most stable clusters at the expense of the population of the less stable ones. 

The final result is a mass spectrum which shows an interesting structure, with peaks of different heights reflecting the abundances of the clusters, related, as indicated above, to their intrinsic stability. In particular, the cluster sizes corresponding to the highest peaks are usually called magic numbers. The specific values of the magic numbers differ in different classes of clusters. For instance,  in clusters of inert gases such as Ne, Ar, Kr and Xe, the magic numbers $N$ = 13, 55, 147, $\ldots$ reflect the packing of weakly interacting spherical atoms forming icosahedra with an increasing number of shells~\cite{EchtPRL1981}. Other well-known example corresponds to clusters of alkaline atoms (Na, K, Rb, Cs), in which the magic numbers $N$ = 8, 20, 40, 58, 92, $\ldots$ reveal the formation of electronic shells by the delocalized valence electrons confined in the mean--field effective potential  well of the cluster~\cite{KnightPRL1984,PedersenNature1991,EkardtPRB1984}.

In addition to the main peaks corresponding to the magic numbers, the mass spectra show a rich structure in between the main peaks. Theoretical modeling and calculations have been of great help to understand the variation of the structure and stability of clusters as a function of size $N$, and to connect this knowledge with the detailed structure of the mass spectra of the families of clusters mentioned above~\cite{FargesJCP1986,EchtJCS1990,SolovyovPRA2002}, as well as other families. Most theoretical calculations use the density functional formalism, because this method has the potential to give accurate predictions for the atomic structure and the binding energy of the clusters. However, a simple comparison of total binding energies, or binding energies per atom, as a function of cluster size is not sensitive enough to reveal accurately the relative stabilities, and more sensitive stability descriptors have been proposed and used. 

Among these, the most popular one is the stability function $\Delta_2(N)$, which measures the total energy of a cluster of size $N$ with respect to the average of the total energies of clusters with sizes $N-1$ and $N+1$. In a plot of $\Delta_2(N)$ as a funtion of $N$, specially stable clusters are characterized by positive values of $\Delta_2(N)$, whilst less stable clusters are characterized by near-zero or negative values of $\Delta_2(N)$. The higher the value of $\Delta_2(N)$, the more stable is the cluster of size $N$ with respect to adjacent clusters with sizes $N+1$ and $N-1$.
 
Nonetheless, some classes of clusters are more complex than inert gas clusters or clusters of simple metals, and $\Delta_2(N)$, although useful, does not explain the full richness of the experimental mass spectra. In addition, the mass spectrum may depend on whether the clusters are born neutral or charged.~\cite{RaoPRB85,JenaCR2018} We propose herein an improved stability index. Since the process of evaporation cooling is important in building the abundance distribution of clusters in the cluster beam, the new stability index $\Delta_2^v(N)$ is derived from the evaporation energies. Specifically, $\Delta_2^v(N)$ is constructed in a way similar  to $\Delta_2(N)$, but with evaporation energies instead of total cluster energies.  We apply this new stability index to understand the experimental abundance of carbon clusters with sizes up to $N=16$ measured by mass spectrometry~\cite{BelauJACS2007}. 

The evolution of the geometrical structure of small carbon clusters is complex~\cite{VanOrdenChemRev1998,RaghavachariJCP1987,JonesPRL1997,JonesJCP1999,CastroJCP2002}, and an accurate description of the structures is required to calculate evaporation energies. In section 2 we briefly present the computational methodology, and give a description of the geometrical, energetic and electronic properties of the different $C_N$ and $C_N^{+}$ ($N\leq$16) gas-phase clusters obtained within our theoretical framework as compared with the broad available experimental and theoretical literature in this field. In Section 3 we apply the most common and extensively used stability descriptors, such as the evaporation energy and the stability function, in order to compare this theoretical analysis with the  experimental mass-spectrometric information on cluster abundances. As a step forward beyond the standard stability descriptors, in Section 4 we propose a new descriptor of the relative stability of clusters that we test by comparing with the time-of-flight mass spectrum of carbon cluster cations $C_N^{+}$ with $N\leq$16. The values of the new stability descriptor as a function of cluster size $N$  provide an accurate correlation with the cluster abundances, reproducing all the details found in the mass spectrum and providing a substantial improvement over the performance of previous descriptors. Finally, we conclude in Section 5 summarizing the conclusions extracted from this study.   

\begin{figure}
\centerline{\includegraphics[width=\columnwidth]{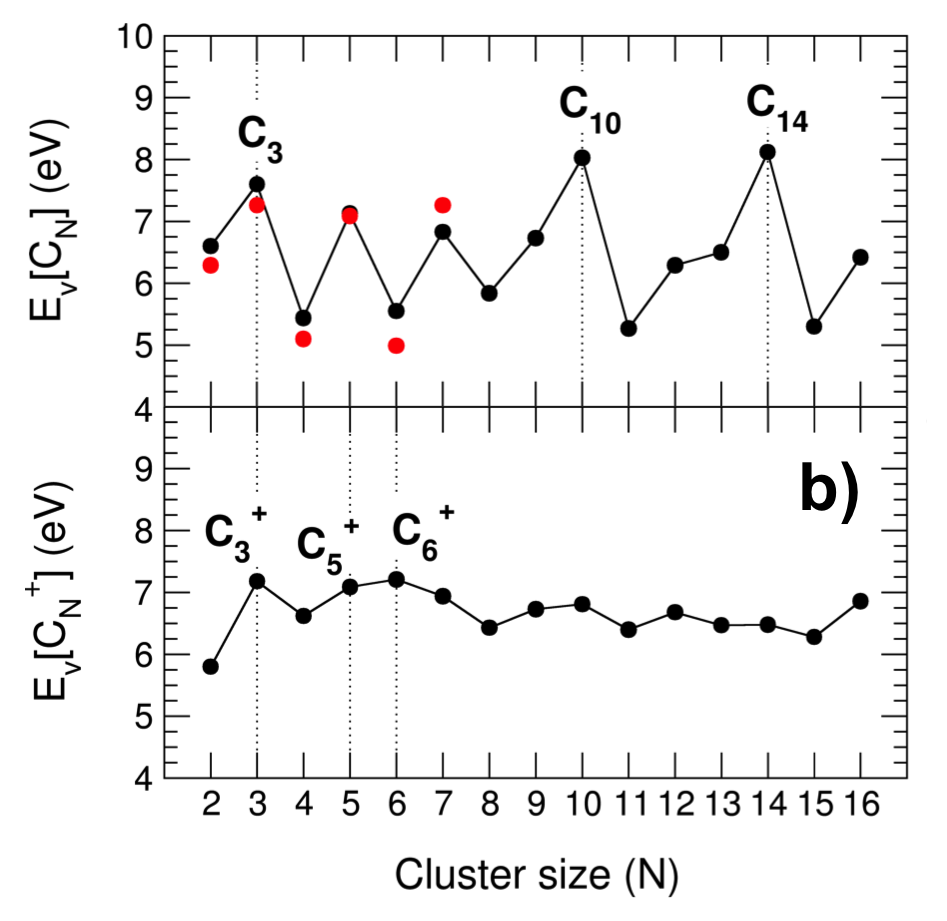}}
\smallskip \caption{Calculated evaporation energies $E_v(N)$, in eV, of: (a) neutral $C_N$ clusters (red dots: experimental values for $C_2$--$C_7$ extracted from Ref.~\cite{GingerichJACS1994}); and (b) positively charged $C_N^{+}$ clusters.
\label{Fig1}} 
\end{figure}

\begin{figure}
\centerline{\includegraphics[width=\columnwidth]{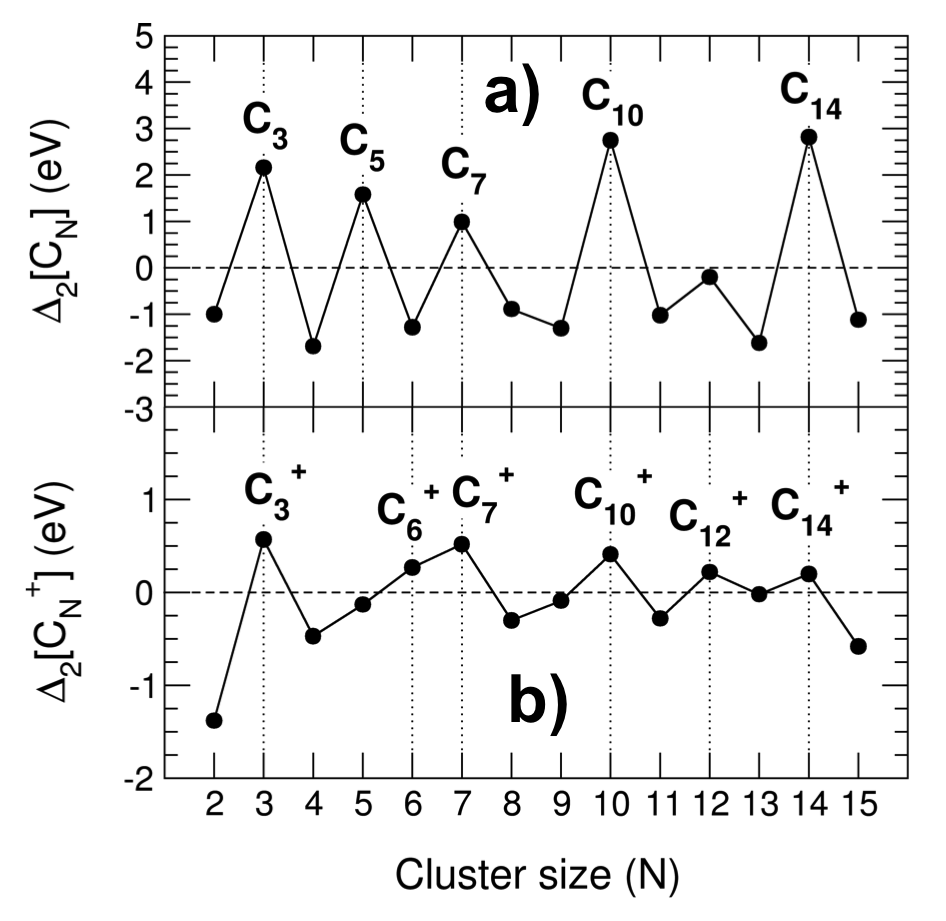}}
\smallskip \caption{Computed stability function $\Delta_2(N)$, in eV, of: (a) neutral $C_N$ clusters; and (b) positively charged $C_N^{+}$ clusters.
\label{Fig3}} 
\end{figure}

\section{Computational Approach and Atomic Structure of $C_N$ and $C_N^{+}$ ($N\leq$16)}

The structure of neutral  $C_N$ and cationic $C_N^{+}$ carbon clusters with $N\leq$16, and the corresponding electronic structures have been optimized by using the {\sc Gaussian09} atomistic simulation package~\cite{g09}. The density functional level employed was the hybrid B3LYP method~\cite{BeckeJCP1993,StephensJPC1994} with cc-PVQZ basis~\cite{DunningJCP1989}. As the starting geometries used as input for the structural optimization we have considered, for each cluster-size, different isomers reported in the literature lying within a total energy window of 0.5 eV. Those isomers had been obtained by accurate molecular orbital calculations (CCSD(T)/cc-pVQZ, CCSD(T)/PVTZ, CCSD(T)/275cGTOs, among others~\cite{VanOrdenChemRev1998}). Besides, for all the isomers studied we have analyzed the two  lowest-lying electronic spin states. Performing a detailed discussion of the geometric and electronic structure of these clusters is not our intention here, given the extensive work already reported in the literature. Our goal herein just consists in carrying out an analysis of the ground-state structures of the $C_N$ and $C_N^{+}$ ($N\leq$16) clusters within the mentioned B3LYP formalism, in order to obtain their global energetic map within an accurate and unified computational framework allowing us to test the new stability descriptor.  

Extensive information exists on  small neutral and cationic carbon clusters obtained from accurate spectroscopic investigations, astronomical observations and theoretical calculations (see, for instance, the excellent review by Van Orden and coworkers~\cite{VanOrdenChemRev1998} and the refereces therein). The ground state and low lying structures that we have obtained for $C_N$ and $C_N^{+}$ ($N\leq$16) clusters with the B3LYP method and cc-PVQZ basis are in excellent agreement with the stablished structures. Our calculations predict that the lowest energy structures of $C_3$, $C_4^{+}$, $C_5$, $C_5^{+}$, $C_7$, $C_7^{+}$, $C_9$ and $C_9^{+}$, besides the obvious $C_2$ and $C_2^{+}$, are cumulenic linear structures with nearly equivalent bond lengths (contrary to acetylenic bonding exhibiting alternating bond lenghts). In agreement with other works the electronic ground states are of type $^{1}\Sigma^{+}_g$($^{3}\Sigma^{-}_g$) for the odd(even)-numbered neutral chains, and $^{2}\Sigma^{+}_u$($^{4}\Sigma^{-}_u$) for the odd(even)-numbered cation chains~\cite{PradhanJCP1994,BrunaJPC1992,Raghavachari1989,RaghavacharJCP1987,WeltnerJACS1971,MartinJCP1990,KurtzAPJ1991,VONHELDEN199433}.   Also in agreement with the reported literature~\cite{VanOrdenChemRev1998,BLEIL1994491,MARTIN1991367,MartinJPC1996,LiangJCP1990}, $C_{11}-C_{16}$ and $C_{11}^{+}-C_{16}^{+}$ clusters exhibit monocyclic-ring structures with  $^{1}A_g$ and $^{2}A^{''}$ electronic ground-states, respectively. This behavior of clusters larger than $C_{10}$ forming monocyclic rings is due to the reduction in angular strain as the radius of the ring increases and to the added stability arising from an additional C--C bond compared to the chains~\cite{RaghavacharJCP1987,MartinJPC1996,LiangJCP1990}. 

Nonetheless, as in previous works we predict a few exceptional cases deviating from the general trend. $C_3^{+}$ which exhibits a bent configuration~\cite{GrevJPC1990,MartinJCP1990-1,ScuseriaCPL1991} ($C_{2v}$ symmetry) with a bond angle of 71$^{\circ}$ (to be compared with an angle of 68$^{\circ}$ reported in literature~\cite{GrevJPC1990}) and a $^{2}B_2$ electronic ground state, 0.19 eV lower in energy that the perfect linear isomer (to be compared with the value of 0.3 eV obtained in an accurate Configuration Interaction (CI) calculation.~\cite{GrevJPC1990}) 

For the neutral $C_4$, $C_6$ and $C_8$, some accurate molecular orbital calculations have found cyclic isomers as low-energy structures, isoenergetic or in some cases even lower in energy than their linear counterparts~\cite{RaghavacharJCP1987,MartinJPC1996,WattsJCP1992,HutterJCP1994,PlessJCP1994,LiangCPL1990}. 
However, these cyclic isomers have been extremely difficult to detect using spectroscopic techniques, while their linear forms habe been observed in most experimental studies. Within our B3LYP/cc-PVQZ calculations we have found the linear cumulenic forms as the most stable structures, with the cyclic isomers lying within an energy range of 0.15 eV. 

For the ground state structure of the cation $C_6^{+}$ we found a slightly non planar distorted hexagon ($D_{3h}$-symmetry) with an $^{2}A^{''}$ electronic ground-state, practically isoenergetic with its linear form. This feature is reproduced by most high-accuracy molecular orbital methodologies, and it has been also confirmed by ion-mobility measurements~\cite{VONHELDEN199433,von_HeldenJCP1991}. The same occurs for the $C_8^{+}$, for which we have found a cyclic ground-state configuration and a low lying linear isomer nearly degenerate in energy, both with $^{2}A^{''}$ electronic ground-states. The structures are again comfirmed by experiment~\cite{VONHELDEN199433}.

In agreement with previous work~\cite{RaghavacharJCP1987}, we found $C_{10}$ as the size for which the transition from linear cumulenic chains to monocyclic rings occurs (allowing for the exceptions discussed above). Our calculation yields  a $D_{5h}$ monocyclic ring as the ground-state configuration, with an $^{1}A_{g}$ electronic ground-state . Finally, we found a cyclic ground state configuration for $C_{10}^{+}$, with a $^{2}A^{''}$  electronic ground-state, again in agreement with previous work~\cite{VONHELDEN199433}.

\section{Stability Descriptors and Correlation with the Mass Spectrum}

\begin{figure}
\centerline{\includegraphics[width=\columnwidth]{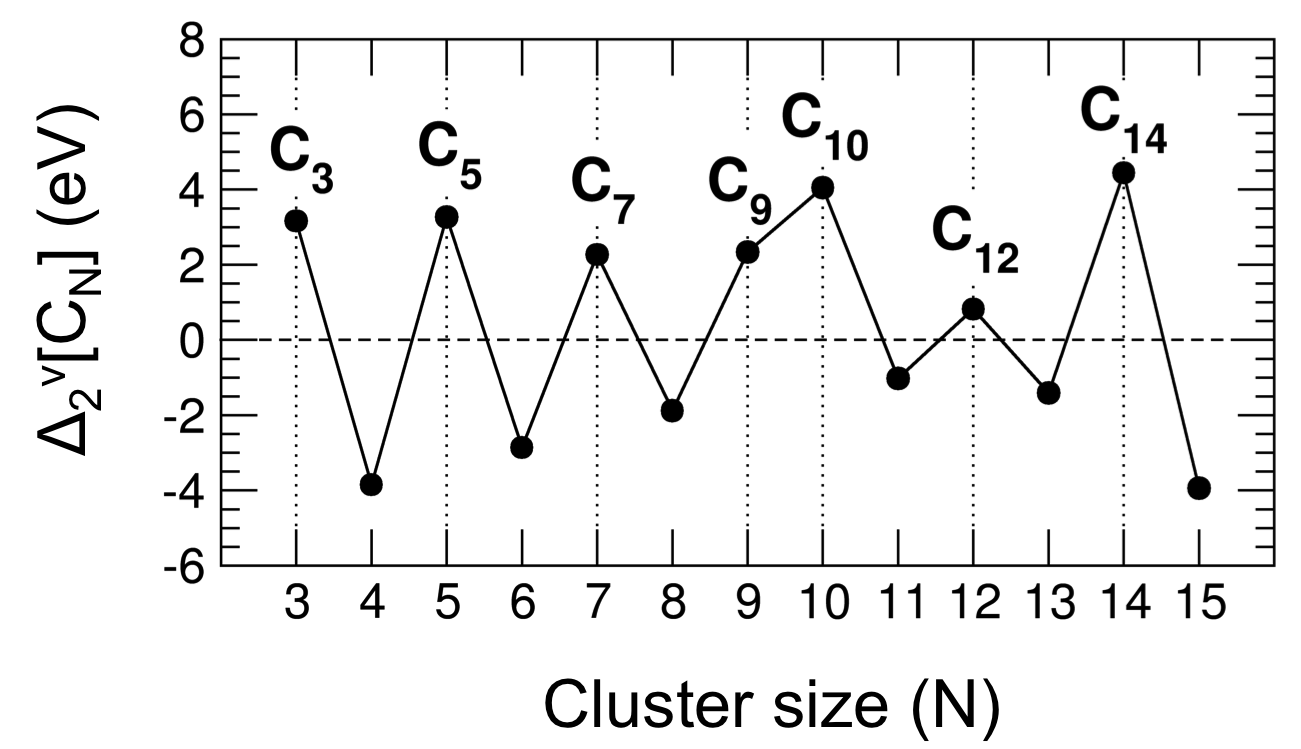}}
\smallskip \caption{Computed stability function, $\Delta_2^v(N)$, (in eV) obtained from evaporation energies of neutral $C_N$ clusters.
\label{Fig3}} 
\end{figure}

The evaporation energy of a carbon cluster with $N$ atoms, $E_v(N)$, is the energy required to detach an atom from the $N$-atom cluster:
\be \label{eq1}
E_v(N)=E(N-1)+E(1)-E(N),
\ee where $E(N)$ is the total energy of an $N$-atom cluster. Correspondingly, $E_v(N + 1)$ is the energy to remove an atom from the cluster $C_{N+1}$:
\be \label{eq2}
E_v(N+1)=E(N)+E(1)-E(N+1).
\ee $E_v(N + 1)$ can also be viewed~\cite{Alonso2003} as a capture energy, $E_c(N)$, that is, the energy released when $C_N$ captures an atom and becomes $C_{N+1}$:
\be \label{eq3}
E_c(N)=E_v(N+1).
\ee

In the case of positively charged carbon clusters $C_{N}^{+}$ the evaporated atom is neutral while the charge remains in the cluster $C_{N-1}^{+}$. The computed evaporation energies of neutral and charged carbon clusters are plotted in Figures 1a and 1b (black dots), respectively. Differences can be observed between the two figures. The largest evaporation energies occur at $C_3$, $C_{10}$ and $C_{14}$ for the neutrals, and at $C_3^{+}$, $C_5^{+}$ and $C_{6}^{+}$ for the cations. Additionally, for sake of comparison, Figure 1a also shows (red dots) the experimental values of the evaporation energies for the neutral clusters $C_2$--$C_7$ extracted from Ref.~\cite{GingerichJACS1994}. The agreement between the computed and experimental values for $C_{N}$ up to $N=$7 is excellent.

From equations~(\ref{eq1}) and (\ref{eq2}) the following relations are immediately obtained:
\bea \label{eq4}
E_v(N)-E_v(N + 1)=E_v(N)-E_c(N)= \\ \nonumber
E(N + 1)+E(N - 1)-2E(N)=\Delta_2(N).
\eea $\Delta_2(N)$ is the ``stability function'' often used to analyze the relative stability of clusters as a function of size, and to interpret the abundance mass spectrum of clusters obtained by gas aggregation techniques. Clusters with large values of $\Delta_2(N)$ are highly stable compared with clusters of neighbor sizes, and the experiments show that those clusters are more abundant. This can be readily understood from equation~(\ref{eq4}). A large value of $\Delta_2(N)$ results from the combination of a large value of $E_v(N)$ and a small value of $E_c(N)$. This means that for that special cluster, both the tendency to grow by capturing one atom and the tendency to shrink the size by evaporating one atom are feeble, so its abundance is enhanced when hot larger clusters evaporate atoms, because that evaporation chain practically stops at the special $C_N$ cluster. 

The quantity given in~(\ref{eq4}) is the atomic parallel of the electronic hardness used in chemistry to quantify the reactivity of a chemical species~\cite{SenBook}. A high chemical hardness is indicative of low chemical reactivity. Here we call $\Delta_2(N)$ of eq.~(\ref{eq4}) the ``atomic hardness''~\cite{Alonso2003} because a high value of this quantity indicates a low predisposition of the cluster to change its size. Figures 2a and 2b show $\Delta_2(N)$ for $C_N$ and $C_N^{+}$ clusters, respectively. The peaks at $C_3$, $C_5$, $C_7$, $C_{10}$ and $C_{14}$ observed for the neutrals indicate that these clusters are more stable than the others. For charged clusters $\Delta_2(N)$ reveals that $C_3^{+}$, $C_6^{+}$, $C_7^{+}$, $C_{10}^{+}$, $C_{12}^{+}$ and $C_{14}^{+}$ are the most stable cationic clusters. 

\begin{figure}
\centerline{\includegraphics[width=\columnwidth]{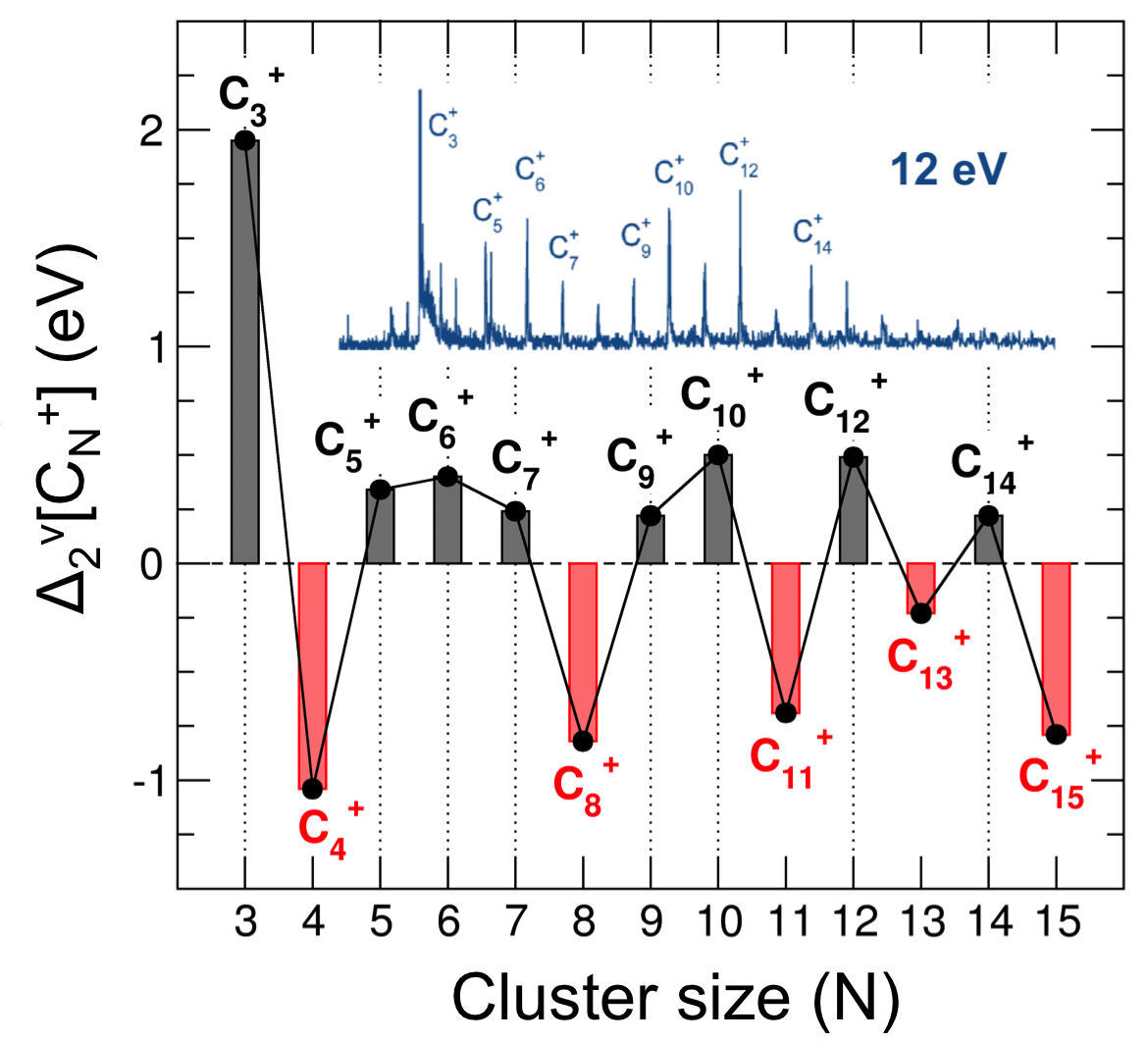}}
\smallskip \caption{Computed stability function, $\Delta_2^v(N)$, (in eV) obtained from evaporation energies of positively charged $C_N^{+}$ clusters. For comparison, the experimental mass spectrum of carbon cluster cations obtained at an ionization energy of 12 eV~\cite{BelauJACS2007} is shown as an inset.
\label{Fig4}} 
\end{figure}

The stability peaks of neutral or cationic clusters do not to show a clear correlation with their corresponding ground-state geometries, since we find a variety of conformations among them: for instance, in the cationic clusters a bent $C_{2v}$ structure for $C_3^{+}$, a distorted $D_{3h}$ hexagon for $C_6^{+}$, a cumulenic linear form for $C_7^{+}$ and monocyclic rings for $C_{10}^{+}$, $C_{12}^{+}$ and $C_{14}^{+}$. It is noticeable that the peak at $N=3$ is not the highest one in the plot for neutral clusters, and the height of the peak at $C_3^{+}$ is similar to that of $C_7^{+}$ and only a bit larger than that of $C_{10}^{+}$. The two stability plots can be compared to the experimental abundance spectrum. Experimental mass spectra obtained by Belau {\it et al.}~\cite{BelauJACS2007} for carbon cluster cations photoionized at 10 and 12 eV yield that, at both ionization energies, the even-numbered clusters in the higher size range ($N$=9--15) are more prominent than the odd-numbered clusters. The most interesting spectrum obtained by Belau and coworkers~\cite{BelauJACS2007} is the one measured at the ionization energy of 12 eV (see inset in Figure 4), sufficiently large to ionize $C_N$ clusters of all the relevant sizes. In fact, at that ionization energy, new peaks are detected, and $C_3^{+}$ becomes the largest peak in the spectrum. In the zone of small clusters, the most abundant ones are $C_3^{+}$, $C_5^{+}$, $C_6^{+}$ and $C_7^{+}$, that is, odd-size clusters, with the addition of $C_6^{+}$. In that plot Belau {\it et al.} mark $C_3^{+}$, $C_5^{+}$, $C_6^{+}$, $C_7^{+}$, $C_9^{+}$, $C_{10}^{+}$, $C_{12}^{+}$ and $C_{14}^{+}$ as abundant clusters. In fact, the peaks at $C_{4}^{+}$, $C_{8}^{+}$, $C_{11}^{+}$, $C_{13}^{+}$ and $C_{15}^{+}$ are clearly small compared to adjacent cations with $N+1$ and $N-1$ atoms. Although the abundance of $C_{5}^{+}$ is smaller than that of $C_{6}^{+}$, the cluster $C_{5}^{+}$ can be considered abundant, and then very stable, because it is not flanked by more abundant neighbors on both sides. The same applies to $C_{7}^{+}$ and $C_{9}^{+}$. Some of the sizes corresponding to abundant clusters are reproduced by the stability functions $\Delta_2(N)$ of Figures 2a and 2b. However, $N$ = 6 and $N$ = 12 are missing in Figure 2a ($\Delta_2(N=12)$ is a negative number) and $N$ = 5 and $N$ = 9 are missing in Figure 2b. Besides the missing peaks, the trend displayed by the relative peak heights in the experimental mass spectrum (inset of Figure 4) is not reproduced well in Figures 2a and 2b. 

\section{New Descriptor of Stability}

We have seen above that $\Delta_2(N)$ provides a rationalization of a number of features seen in the experimental spectrum, but not all features. With the aim of improving the theoretical description and the prediction of all the features in the experimental spectrum, herein we introduce a new stability descriptor. For that purpose we write the evaporation energies for clusters with $N + 1$, $N$ and $N - 1$ atoms:
\be \label{eq5}
E_v(N + 1) = E(N) + E(1) - E(N + 1),
\ee
\be \label{eq6}
E_v(N) = E(N - 1) + E(1) - E(N),
\ee
\be \label{eq7}
E_v(N - 1) = E(N - 2) + E(1) - E(N - 1).
\ee
From these equations:
\bea \label{eq8}
E_v(N) - E_v(N+1) &=& \\ \nonumber
E(N+1) + E(N-1) - 2E(N) &=& \Delta_2(N).
\eea and
\bea \label{eq9}
E_v(N-1) - E_v(N) &=&  \\ \nonumber
E(N) + E(N-2) - 2E(N-1) &=& \Delta_2(N-1).
\eea
Subtracting~(\ref{eq9}) from~(\ref{eq8}), we obtain $2E_v(N) - E_v(N + 1) - E_v(N - 1)$, which, by comparing with eq.~(\ref{eq4}), is an stability descriptor defined from the evaporation energies. That is,
\bea \label{eq10}
\Delta_2^v(N) = 2E_v(N) - E_v(N + 1) - E_v(N -1) = \\ \nonumber
\Delta_2(N) - \Delta_2(N-1),
\eea
which is the difference between the ``atomic hardness'' of the clusters with $N$ and $N-1$ atoms. One may notice the change in the relative order of the quantities pertaining to $N$, $N+1$ and $N-1$ compared to eq.~(\ref{eq4}). The reason is that total energies $E(N)$ are negative, but evaporation energies $E_v(N)$ are positive. The interpretation of eq.~(\ref{eq10}) is that a cluster of size $N$ is very stable if its evaporation energy is larger than the average of the evaporation energies of the clusters with adjacent sizes $N+1$ and $N-1$. Another way to interpret eq.~(\ref{eq10}) is that a cluster is very stable, and abundant, if its atomic hardness is high and the hardness of clusters with size $N-1$ is low. In this way, the tendencies of clusters of size $N$ to grow or to shrink are feeble, and at the same time the population of clusters of size $N$ can easily grow in the cluster beam when clusters of size $N-1$ capture one atom. Very stable clusters are characterized by positive values of $\Delta_2^v(N)$. Results obtained with this descriptor of cluster stability for neutral and cationic clusters are plotted in Figures 3 and 4, respectively. An improvement is noticed in Figure 3 compared to the results for the standard stability index $\Delta_2(N)$. Now the only missing feature in comparison to the experimental spectrum is the peak at $C_6$. However, the relative heights of the peaks do not correlate with the heights of the peaks in the experimental mass spectum. This suggests the need of accounting for the charge of the cluster and Figure 4 confirms this expectation. $\Delta_2^v(N)$ separates the cations in two groups:   $C_3^{+}$, $C_5^{+}$, $C_6^{+}$, $C_7^{+}$, $C_9^{+}$, $C_{10}^{+}$, $C_{12}^{+}$ and $C_{14}^{+}$ have positive values of $\Delta_2^v(N)$, and $C_{4}^{+}$, $C_{8}^{+}$, $C_{11}^{+}$, $C_{13}^{+}$ and $C_{15}^{+}$ have negative values of $\Delta_2^v(N)$, indicating that the first group is formed by the more stable (and then more abundant) cationic clusters, in full agreement with experiment. A remarkable observation is the correlation between the heights of the peaks in Figure 4 and the corresponding heights in the experimental spectrum. $C_{3}^{+}$ appears as the largest peak in the spectrum. Also, the trends in the two groups ($C_5^{+}$,$C_6^{+}$,$C_7^{+}$) and ($C_9^{+}$,$C_{10}^{+}$,$C_{12}^{+}$,$C_{14}^{+}$) are well reproduced.

To justify the clear improvement obtained by using the new stability index $\Delta_2^v(N)$ instead of  $\Delta_2(N)$  we notice from eq.~(\ref{eq4}) that $\Delta_2(N)$ is constructed from the total energies of clusters with sizes $N$+1, $N$ and $N$-1. On the other hand $\Delta_2^v(N)$ is constructed from evaporation energies of clusters with sizes $N$+1, $N$ and $N$-1. Evaporation energies are more relevant quantities in the processes controlling the relative cluster abundances in the beam, so it is reasonable that $\Delta_2^v(N)$ provides a better correlation with the mass spectrum of carbon cluster cations.

\section{Summary and Conclusions}
The abundance mass spectrum of gas-phase atomic clusters measured in cluster beams has a rich structure that arises from a complex process of growth by atom addition and evaporation when the clusters become hot due to the heat of condensation. The most popular stability descriptor, usually called "stability function", is based on comparing the total energy of clusters of size $N$ with the averaged total energy of clusters with neighbor sizes $N$+1 and $N$-1. Those total energies have to be obtained by laborious calculations of the atomic and electronic structures of the clusters using powerful theoretical methods, usually the density functional theory. This index explains the main features of the experimental mass spectra, but application to carbon cluster cations $C_N^{+}$ ($N\leq$16) leaves some secondary features, and in particular the detailed behavior of the height of the experimental abundance peaks, unexplained.  In this work we have introduced a novel stability descriptor which is based on comparing the evaporation energy of clusters of size $N$ with the averaged evaporation energies of clusters of sizes $N$+1 and $N$-1. Application to carbon cluster cations $C_N^{+}$ ($N\leq$16) reveals a perfect correlation with the experimental mass spectrum, and in particular with the detailed variation of the height of the abundance peaks as a function of size $N$. The success is attributed to the fact that the normal stability index is constructed from total cluster energies, while the new index is constructed from atom--evaporation energies, which are quantities more directly connected with the processes controlling the abundances in the beam. We expect that the new stability index will be useful in the analysis of the mass spectra of all kinds of clusters. The index is based on the energies involved in the processes of evaporation and capture of atoms by the clusters forming and flying in the molecular beam. A part of the condensation energy of the clusters is removed by collisions with a cool inert carrier gas. Another part is removed by evaporation of atoms from the hot clusters. For the detailed variations of the abundance population with cluster size to develop, abundant evaporation events are required, and this is usually the case in cluster formation experiments in the gas phase.

\section*{Acknowledgements}
We acknowledge funding from the Spanish MINECO (Grants MAT2017-85089-C2-1-R and RYC-2015-17730), Junta de Castilla y Le\'on (Grant VA021G18), University of Valladolid (Grupo de F\'isica de Nanoestructuras), and the EU via the ERC-Synergy Program (Grant ERC-2013-SYG-610256 {\sc Nanocosmos}) and EU Horizon 2020 Research and Innovation program (Grants 696656 -- Graphene Flagship Core1 -- and 785219 -- Graphene Flagship Core2 --). We also thank the computing resources from CTI-CSIC.

\

The authors declare no competing financial interests.

\bibliography{bib} 

\end{document}